\documentclass[%
    reprint,
    amsmath,
    amssymb,
    aps，
]{revtex4-2}

\usepackage{graphicx}
\usepackage{dcolumn}
\usepackage{bm}

\begin{document}

\title{Absence of Weyl nodes in EuCd$_2$As$_2$ revealed by the carrier density dependence of the anomalous Hall effect}

\author{Yue Shi$^{1,2}$}
\author{Zhaoyu Liu$^{1}$}
\author{Logan A. Burnett$^{3}$}
\author{Seokhyeong Lee$^{4}$}
\author{Chaowei Hu$^{1}$}
\author{Qianni Jiang$^{1}$}
\author{Jiaqi Cai$^{1}$}
\author{Xiaodong Xu$^{1,2}$}
\author{Mo Li$^{1,4}$}
\author{Cheng-Chien Chen$^{3}$}
\author{Jiun-Haw Chu$^{1}$}

\affiliation{$^{1}$Department of Physics, University of Washington, Seattle, Washington, 98195, USA}
\affiliation{$^{2}$Department of Materials Science and Engineering, University of Washington, Seattle, Washington, 98195, USA}
\affiliation{$^{3}$Department of Physics, University of Alabama at Birmingham, Birmingham, Alabama, 35294, USA}
\affiliation{$^{4}$Department of Electrical and Computer Engineering, University of Washington, Seattle, Washington, 98195, USA}

\date{\today}

\begin{abstract}
The antiferromagnetic layered compound EuCd$_2$As$_2$ is widely considered as a leading candidate of ideal Weyl semimetal, featuring a single pair of Weyl nodes in its field-induced ferromagnetic (FM) state\cite{4_Ideal_WSM,5_LLWang_singleWFermions,10_induced_WSM}. Nevertheless, this view has recently been challenged by an optical spectroscopy study, which suggests that it is a magnetic semiconductor\cite{12_ECA_semiconductor}. In this study, we have successfully synthesized highly insulating EuCd$_2$As$_2$ crystals with carrier density reaching as low as $2\times 10^{15}$ $\text{cm}^{-3}$. The magneto-transport measurements revealed a progressive decrease of the anomalous Hall conductivity (AHC) by several orders of magnitude as the carrier density decreases. This behavior contradicts with what is expected from the intrinsic AHC generated by the Weyl points, which is independent of carrier density as the Fermi level approaches the charge neutrality point. In contrast, the scaling relationship between AHC and longitudinal conductivity aligns with the characteristics of variable range hopping insulators. Our results suggest that EuCd$_2$As$_2$ is a magnetic semiconductor rather than a topological Weyl semimetal.
\end{abstract}

\maketitle

\section{\label{sec:level1}Introduction}

\begin{figure*}[th]
\includegraphics[width = 6.8 in]{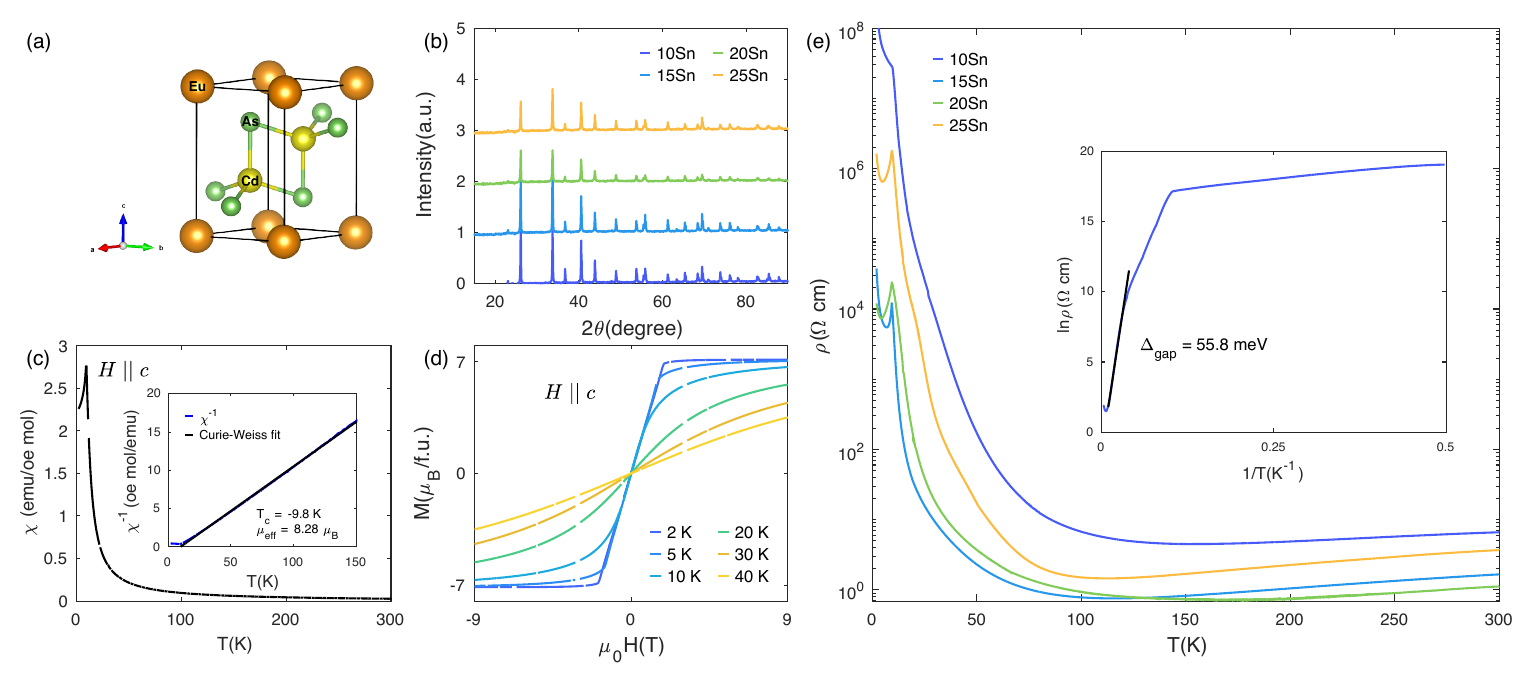} 
\caption{\textbf{Transport and magnetic characterizations of EuCd$_2$As$_2$.}
\textbf{(a)} Crystal structure of EuCd$_2$As$_2$ with P-3m1 space group. \textbf{(b)} Powder X-ray diffraction of EuCd$_2$As$_2$ at room temperature. 
\textbf{(c)} Temperature dependence of magnetic susceptibility $\chi$ with 0.1 T magnetic field applied along the $c$-axis. The inset shows the inverse susceptibility versus temperature and the Curie-Weiss fitting. \textbf{(d)} Magnetization versus field at different temperatures. The saturated magnetization is $\sim$ 7 $\mu_{B}$ per Eu atom at the base temperature 2 K.
\textbf{(e)} Temperature dependence of the electrical resistivity $\rho(\text{T})$ between 2-300 K. The inset shows $\ln\rho$ versus the inverse of temperature $1/\text{T}$ of the sample from 10Sn. The transport activation gap $\Delta_{gap} =$ 55.8 meV is obtained by fitting $\ln\rho$ versus $1/\text{T}$ in the temperature range of 30-100K.}
\label{fig1}
\end{figure*}

Weyl semimetals have attracted significant attention due to their intriguing electronic properties arising from the presence of the Weyl points\cite{WSM_review_Armitage,WSM_review_Yan}. Among them, the ideal magnetic Weyl semimetals, characterized by a single pair of Weyl points close to the Fermi level, are long sought for exploring the intrinsic topological effects associated with Weyl fermions free from interference by other trivial energy bands. In the past few years, the layered magnetic compound EuCd$_2$As$_2$ was proposed to be one of the leading candidates to realize ideal magnetic Weyl semimetal in its field-induced ferromagnetic (FM) state\cite{4_Ideal_WSM, 5_LLWang_singleWFermions, 10_induced_WSM}. EuCd$_2$As$_2$ has a hexagonal unit cell with space group P-3m1 (No.164) (Fig. \ref{fig1}(a)). At temperature of 9.5 K, the localized magnetic moments of Eu develop a long-range magnetic order, with moments ferromagnetically coupled within layers and antiferromagnetically coupled between layers\cite{Wang2016, 6_magnetic_order, 7_spectroscopic_of_EuCd2X2}. Application of a 2 Tesla magnetic field saturates all magnetic moments along the $c$-axis at the low temperature, stabilizing a field-induced FM state. Previous density functional theory (DFT) calculations suggested that Cd $s$ and As $p$ orbitals are inverted near the $\Gamma$ point in the energy bands of EuCd$_2$As$_2$\cite{8_type-IV_magnetic_space_groups,6_magnetic_order}, leading to an antiferromagnetic (AFM) topological insulator. Once the moments are fully saturated along the $c$-axis in the FM state, the Zeeman splitting lifts the band degeneracy and gives rise to a single pair of Weyl points\cite{5_LLWang_singleWFermions, 10_induced_WSM, 4_Ideal_WSM}. 

In previous works, high magnetic field quantum oscillation study has revealed a small Fermi surface with a light effective mass in the field-induced FM state of EuCd$_2$As$_2$\cite{4_Ideal_WSM}; Hall resistivity study has presented a large anomalous Hall conductivity at low temperatures presumably related with topological state of EuCd$_2$As$_2$\cite{Xu2021UnconventionalTransverse,Cao2022nonlinearAHE,9_AHC,4_Ideal_WSM}; ARPES experiments observed the quasi-linear dispersion of the valence band in the AFM state and the band degeneracy splitting at temperature right above the Neel temperature due to the existence of strong spin fluctuations\cite{Wang2022ARPES,11_Dirac_AFM_ECA,10_induced_WSM}. All of these observations seem to agree with the existence of Weyl semimetal phase. Nevertheless, the direct observation of Weyl crossings of EuCd$_2$As$_2$ is not possible because all the samples synthesized so far are $p$-type with Fermi level below the valence band top. In contrast, a recent study employing optical conductivity and pump-probe photoemission has suggested that EuCd$_2$As$_2$ is, in fact, a semiconductor with an electronic gap of 0.77 eV instead of a Weyl semimetal\cite{12_ECA_semiconductor}. Furthermore, a systematic investigation on the DFT calculation of materials family EuCd$_2$X$_2$ (X=P, As, Sb, Bi) revealed that the topological states of materials in this family are sensitive to the choice of functional and the value of Hubbard \textbf{U}, hence the conduction and valence band might not even be inverted in EuCd$_2$As$_2$\cite{PhysRevB.108.075150}.

A well known prediction for a type-I Weyl semimetal is that the anomalous Hall conductivity (AHC) is solely determined by the distance between the Weyl nodes in the momentum space when the Fermi level is exactly at the Weyl crossings\cite{Yang2011,Burkov2014}. Therefore, the AHC is expected to approach a finite and constant value as carrier density decreases to zero, which serves as a direct evidence to distinguish Weyl semimetals from other trivial electronic states. In this work, we successfully synthesized EuCd$_2$As$_2$ single crystals with carrier density ranging from $10^{15}$ to $10^{16}$ $\text{cm}^{-3}$ in the field induced FM state and conducted a systematic study of the transport behaviors of these samples. We found that the temperature dependence of resistivity of all samples show an insulating behavior, and the resistivity dramatically decreases by several orders of magnitude as Eu moments are fully saturated by a $c$-axis magnetic field. We found the AHC decreases by several orders of magnitude as the carrier density decreases, far below the value expected from previous DFT calculations that predicted Weyl semimetal phase in this system. In contrary, scaling analysis shows good agreement with the AHC in the variable range hopping regime. The optical transmittance measurement further confirms a finite optical gap of 0.74 eV, consistent with previous study. \textit{Ab initio} calculation to compute the Hubbard \textbf{U} parameter of EuCd$_2$As$_2$ was also performed in this work, which reveals that previous calculations significantly underestimate the value of Hubbard \textbf{U}, highlighting the importance of correlation when determining the topological phase of magnetic semiconductors.

\section{\label{sec:level1}Results}
\subsection{\label{sec:level2}Transport and Magnetic Characterization}

\begin{figure*}[t]
\includegraphics[width = 5.8 in]{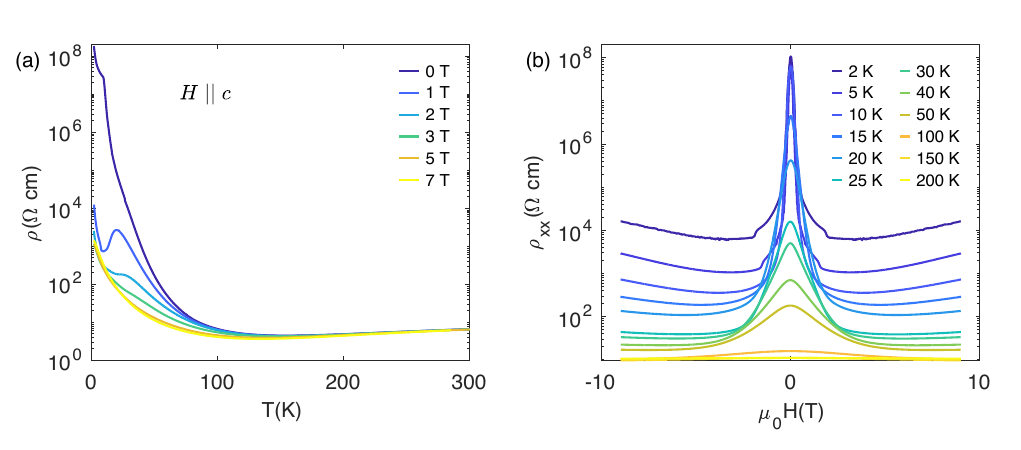} 
\caption{\textbf{Colossal magnetoresistance of EuCd$_2$As$_2$.}
\textbf{(a)} Temperature dependence of resistivity, $\rho(\text{T})$, under different magnetic fields applied along the $c$-axis. \textbf{(b)} Resistivity $\rho_{xx}$ versus magnetic field at temperatures between 2-200 K.}
\label{fig2}
\end{figure*}

EuCd$_2$As$_2$ single crystals were synthesized via the Sn flux method\cite{13_Manipulate_magnetic_inECA}. We discovered that the residual carrier density of crystals is sensitive to the Sn flux ratio. The stoichiometric Eu pieces, Cd and As shots are mixed with 10 to 25 times atomic mass Sn shots as starting materials, which yielded EuCd$_2$As$_2$ crystals with hole carrier density spanning from $10^{15}$ to $10^{16}$ $\text{cm}^{-3}$. While it is unclear how the Sn flux ratio alters the defect density, the wide range of carrier density offers us a convenient knob to tune the transport properties of EuCd$_2$As$_2$.

Figure \ref{fig1}(b) shows the powder X-ray diffraction of EuCd$_2$As$_2$ samples from batches using 10, 15, 20, and 25 Sn flux ratio (labeled as 10Sn, 15Sn, 20Sn, and 25Sn). All diffraction patterns and extracted lattice constants are identical to the previous report\cite{13_Manipulate_magnetic_inECA}, indicating that the change of Sn flux ratio has minimal effect on structural properties. Similarly, the magnetic properties of samples from different batches show no measurable difference and agree with previous studies\cite{10_induced_WSM}. The representative data of a sample from 10Sn is summarized in Fig. \ref{fig1}. As shown in Fig. \ref{fig1}(c), the temperature dependence of $c$-axis magnetic susceptibility $\chi$ follows the Curie-Weiss temperature dependence, with a kink at 9.5 K indicating the onset of AFM order. Fitting the high temperature susceptibility with the Curie-Weiss law yields a Weiss temperature of $-$9.8 K and effective magnetic moment of 8.28 $\mu_{B}$. Figure \ref{fig1}(d) shows the magnetization $\textbf{M}$ as a function of the field measured at different temperatures. At the base temperature 2 K. $\textbf{M}$ increases linearly before reaching the saturated value of $\sim$ 7 $\mu_{B}$ per $\text{Eu}$ atom, in agreement with the magnetic moment of a divalent Eu with half-filled 4$f$ orbitals.

Unlike the structural and magnetic properties, the electrical transport of EuCd$_2$As$_2$ is very sensitive to the Sn flux ratio. Figure \ref{fig1}(e) shows the temperature dependence of in-plane resistivity $\rho(\text{T})$ for samples selected from different batches. All samples show an insulating behavior below 100 K, where resistivity increases by three to seven orders of magnitude as the temperature decreases. This is very different from the semimetal behavior reported in previous literatures\cite{4_Ideal_WSM,Xu2021UnconventionalTransverse,Cao2022nonlinearAHE}, with the exception of the most recent works\cite{12_ECA_semiconductor,wang2023absence}. The kink of resistivity at 9.5 K is consistent with the AFM transition determined from the magnetic susceptibility. Below 9.5 K the resistivity of the sample from 10Sn continuously increases, while samples from 15Sn, 20Sn, and 25Sn initially decrease before increasing again as the temperature cools further down to 2 K. The inset of Fig. \ref{fig1}(e) shows $\ln\rho$ versus $1/\text{T}$ of the sample from 10Sn. The black solid line denotes the linear fit of the $\ln\rho$ versus $1/\text{T}$ in the temperature range between 30 to 100 K, which allows us to extract the activation energy $\Delta_{gap} \sim$ 55.8 meV using the relationship $\ln\rho = \frac{\Delta_{gap}}{2K_BT}$. We note that in a disordered semiconductor the activation energy extracted from resistivity is often related to the energy of impurity levels, which is much lower than the bandgap. 

\subsection{\label{sec:level2}Colossal Magnetoresistance}

The insulating resistivity of EuCd$_2$As$_2$ can be strongly modulated by the application of magnetic fields. The magnetoresistance of the sample from 10Sn is summarized in Fig. \ref{fig2}. The results are qualitatively similar to samples from other batches. Figure \ref{fig2}(a) shows the temperature dependence of the resistivity $\rho(\text{T})$ under different magnetic fields from 0 to 7 T. The exponential increase of resistivity is suppressed by the external magnetic field, resulting in a dramatic decrease of resistivity by almost five orders of magnitude.  As the magnetic field increases, the kink at Neel temperature is smeared, and a bump-like feature at $\sim$ 20 K emerges for 1 T magnetic field and then shifts to the higher temperatures. For temperatures higher than 100 K, where resistivity is no longer following the thermally activated temperature dependence, the magnetoresistance becomes negligible. 

Figure \ref{fig2}(b) shows the magneto-resistivity $\rho_{xx}$ as a function of field at a fixed temperature between 2-200 K. It can be seen that at low temperatures resistivity abruptly drops by several orders of magnitude as magnetic field increases, followed by a kink at the saturation field, above which the resistivity increases again when magnetic moments are fully aligned with the field. As temperature increases, the drop of resistivity becomes more gradual and the kink is smeared out, which is highly correlated with the field dependence of magnetization at higher temperatures (Fig. \ref{fig1}(d)). This is also reported in a recent work\cite{wang2023absence}, and consistent with the decrease of bandgap induced by the saturation of magnetization reported in Ref.\cite{12_ECA_semiconductor}. The $\text{MR}=100\%\times(\rho_{xx}(9 \text{T})-\rho_{xx}(0 \text{T}))/\rho_{xx}(0 \text{T})$ reaches negative 90\% at $T<50$ K. Such a colossal negative MR was also observed in several Eu-based semiconducting materials, a property that might be of interests for spintronics applications\cite{14_Colossal_MR_Rev, 15_EuP3}. 

\begin{figure*}[t]
\includegraphics[width= 5.8 in]{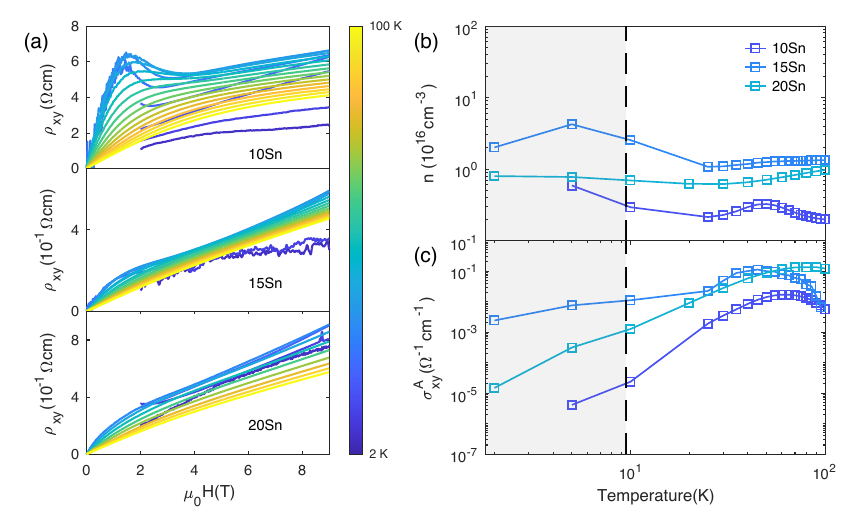} 
\caption{\textbf{The temperature dependence of the carrier density and anomalous Hall conductivity of EuCd$_2$As$_2$.}
\textbf{(a)} Hall resistivity $\rho_{xy}$ versus magnetic field for samples from 10Sn, 15Sn and 20Sn, at temperatures between 2-100 K. \textbf{(b)} Carrier density $n$, and \textbf{(c)} Anomalous Hall conductivity $\sigma_{xy}^{A}$ versus temperature obtained by fitting from $\rho_{xy}$ in left panel (see main text for details of the fitting). Vertical lines mark antiferromagnetic transition temperature $T_N$ = 9.5 K.}
\label{fig3}
\end{figure*}

\begin{figure*}[t]
\includegraphics[width = 6 in]{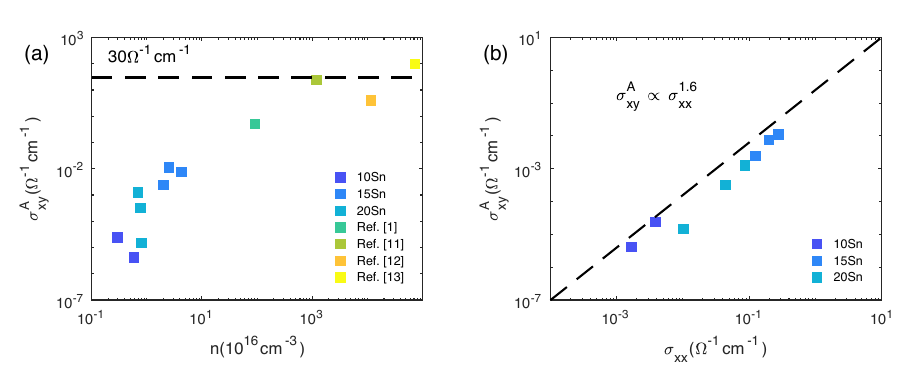} 
\caption{\textbf{Carrier density dependence and the scaling relationship of the anomalous Hall effect in EuCd$_2$As$_2$.}
\textbf{(a)} Anomalous Hall conductivity $\sigma_{xy}^{A}$ versus carrier density $n$ measured from samples 10Sn, 15Sn, 20Sn, and semi-metallic EuCd$_2$As$_2$ reported in previous works\cite{4_Ideal_WSM,Xu2021UnconventionalTransverse,Cao2022nonlinearAHE,9_AHC}. The expected value of $\sigma_{xy}^{A}$ (black dash line) for an ideal Weyl semimetal with only one pair of Weyl nodes separated by 0.52 $\text{nm}^{-1}$\cite{4_Ideal_WSM}.
\textbf{(b)} Anomalous Hall conductivity $\sigma_{xy}^{A}$ versus longitudinal conductivity $\sigma_{xx}$ in log-log scale. The dash line marks the power law $\sigma_{xy}^{A}\propto\sigma_{xx}^{1.6}$.}
\label{fig4}
\end{figure*}

\subsection{\label{sec:level2}Anomalous Hall Effect}
The Hall resistivity $\rho_{xy}$ are shown in Fig. \ref{fig3}(a) for samples from 10Sn, 15Sn and 20Sn. Similar to the longitudinal resistivity $\rho_{xx}$, $\rho_{xy}$ is also strongly affected by the magnetism. As temperature decreases, the field dependence of $\rho_{xy}$ becomes highly non-linear and resembles that of the magnetization, a clear sign of anomalous Hall effect (AHE). For temperatures below 10 K, $\rho_{xx}$ becomes too large ($\sim$ M $\Omega\text{cm}$) for magnetic fields below 2 T so that anti-symmetrization becomes too difficult to extract $\rho_{xy}$, and hence only data above 2 T is presented. To quantitatively understand the temperature and carrier density dependence of AHE, we extract the anomalous Hall resistivity, $\rho_{xy}^{A}$, and the ordinary Hall effect, $R_{0}\mu_0$H, from the linear fit of $\rho_{xy}$ in the field range of 3 to 9 T using the expression $\rho_{xy}=R_{0}\mu_0 H +\rho_{xy}^{A}$. The ordinary Hall coefficient, $R_{0}$, allows us to calculate the carrier density $n$. The anomalous Hall resistivity $\rho_{xy}^{A}$ is converted to AHC $\sigma_{xy}^{A}$ via the expression $\sigma_{xy}^{A}=\rho_{xy}^{A}/({\rho_{xy}^{A}}^2+\rho_{xx}^{2})$, where the values of $\rho_{xx}$ is taken at 9 T, at which the magnetic moments are most aligned along the $c$-axis. Using $\rho_{xx}$ at different fields results in minor changes of the absolute values AHC, but does not change its dependence on carrier density or longitudinal conductivity. Figures \ref{fig3}(b) and \ref{fig3}(c) show the temperature dependence of carrier density $n$ and AHC $\sigma_{xy}^{A}$ between 2 and 100 K. Intriguingly, the carrier densities of all three samples show a non-monotonic temperature dependence, which is in contrast to the thermally activated behavior where carrier density decreases monotonically with temperature. This is likely due to the interplay between the thermal activation of carriers and the change of magnetic configurations at elevated temperatures due to moment fluctuations. Regardless of the origin, the carrier density and the AHC measured across three samples show a substantial variation for $T < T_N$, providing us an opportunity to systematically investigate the carrier density dependence of AHC of EuCd$_2$As$_2$ in the field induced FM state. 

In a Weyl semimetal, the Weyl nodes are the source and sink of Berry curvature flux in the momentum space. In an ideal time-reversal symmetry breaking Weyl semimetal, the single pair of Weyl nodes form a Berry curvature dipole gives rise to an intrinsic AHE solely depending on the separation of the Weyl nodes, $\sigma_{xy}^{A}=(e^{2}/2\pi h\Delta k)$. This $\sigma_{xy}^{A}$ is topological and independent of the Fermi level position, hence we should expect a constant AHC even if the carrier density approaches zero. Here, we compiled the data of $\sigma_{xy}^{A}$ and $n$ extracted below 10 K in the field range of 3-9 T, when EuCd$_2$As$_2$ is in the field induced FM state, from Figs. \ref{fig3}(b) and \ref{fig3}(c), and plotted in Fig. \ref{fig4}(a) in a log-log scale. In addition to the data collected in this work (blue squares), the AHC measured from higher carrier density samples in the previous works are also included\cite{4_Ideal_WSM,Xu2021UnconventionalTransverse,Cao2022nonlinearAHE,9_AHC}. The horizontal black dash line indicates the value of the intrinsic AHC, $\sim30$ $\Omega^{-1}{\text{cm}}^{-1}$, from a single pair of Weyl points separated by 0.52 $\text{nm}^{-1}$ in ferromagnetic EuCd$_2$As$_2$ predicted by previous DFT calculations\cite{4_Ideal_WSM, 5_LLWang_singleWFermions}. As shown in Fig. \ref{fig4}(a), the observed $\sigma_{xy}^{A}$ has a strong dependence of carrier density $n$ and decreases dramatically as $n$ decreases. As the carrier density approaches $\sim10^{15}$ $\text{cm}^{-3}$, the AHC reaches a value that is five orders of magnitude smaller than the AHC expected for an ideal Weyl semimetal.

To gain insight into the underlying mechanism of the AHE in EuCd$_2$As$_2$, the $\sigma_{xy}^{A}$ versus $\sigma_{xx}$ ($\sigma_{xx} = \rho_{xx}/(\rho_{xy}^{2}+\rho_{xx}^{2})$) are plotted in Fig. \ref{fig4}(b). Note that we use the value of resistivity tensors at $\mu_0 H=9$ T for all conversions since both longitudinal and Hall resistivities depend strongly on the magnetic fields. As shown in the figure, all longitudinal conductivity $\sigma_{xx}$ fall below the value of 1 $\Omega^{-1}\text{cm}^{-1}$, which belongs to the insulating regime, where conduction and AHE are governed by the variable range hopping or activated hopping process. Moreover, the AHC versus longitudinal conductivity can be well fitted by a power law relation, $\sigma_{xy}^{A}\propto\sigma_{xx}^{\alpha}$, with $\alpha=1.6$ (black dash line). This is consistent with the previous experimental results and theoretical models of AHE in the insulating regime\cite{AHE_RMP}. Thus, the observed AHC is fully consistent with EuCd$_2$As$_2$ being a magnetic semiconductor.

\subsection{\label{sec:level2}Optical Transmittance}

\begin{figure}[h]
\includegraphics[width = 3.2 in]{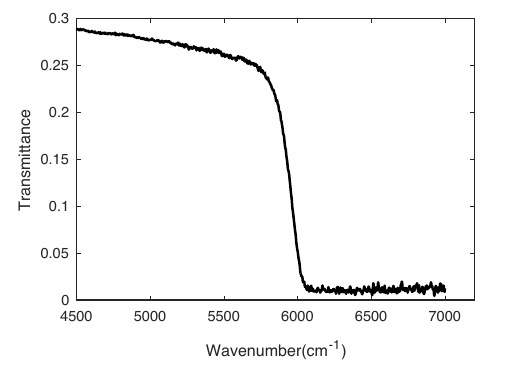} 
\caption{\textbf{Transmittance spectrum of EuCd$_2$As$_2$.}
Transmittance versus incident light wavenumber of sample 10Sn measured by FTIR at room temperature. The incident lights with wavenumber larger than 6000 $\text{cm}^{-1}$ are fully absorbed, consistent with a bandgap of 0.74 eV.}
\label{fig5}
\end{figure}

As a final confirmation, the optical transmittance spectrum of sample 10Sn was measured using the Fourier Transform Infrared spectroscopy (FTIR) technique at room temperature. As shown in Fig. \ref{fig5}, EuCd$_2$As$_2$ remains transparent up to a wavenumber of 6000 $\text{cm}^{-1}$(equivalent to $\sim$ 0.74 eV), followed by a sharp drop of transmittance at higher wavenumber. This is consistent with the presence of a $\sim$ 0.74 eV bandgap, which is in agreement with the optical study reported in Ref.\cite{12_ECA_semiconductor}. 

\section{\label{sec:level1}Discussion}
The comprehensive anomalous Hall effect and optical transmittance measurements presented above clearly support the conclusion that EuCd$_2$As$_2$ is a magnetic semiconductor. A remaining question is why most of the previous first-principles calculations identify ferromagnetic EuCd$_2$As$_2$ as a Weyl semimetal. We note that recent calculations by Cuono \textit{et al.}\cite{PhysRevB.71.035105} have shown that electron correlation effects can significantly alter the electronic and topological behaviors of EuCd$_2$As$_2$. In particular, within the standard generalized gradient approximation (GGA) functional, antiferromagnetic (AFM) EuCd$_2$As$_2$ can exhibit a vanishing electronic band gap and nontrivial band topology only when Eu’s on-site Hubbard repulsion \textbf{U} is less than $\sim$ 6 eV\cite{PhysRevB.108.075150}. Previous theoretical and experimental studies have considered \textbf{U} values in the range of $\sim$ 3-5 eV\cite{8_type-IV_magnetic_space_groups,11_Dirac_AFM_ECA,10_induced_WSM,PhysRevB.98.125110,6_magnetic_order}, which could tend to conclude that EuCd$_2$As$_2$ is topologically nontrivial. Since the Hubbard repulsion plays a critical role, it is important to more accurately estimate the \textbf{U} value.

\begin{figure}[ht!]
\includegraphics[width = 2.8 in]{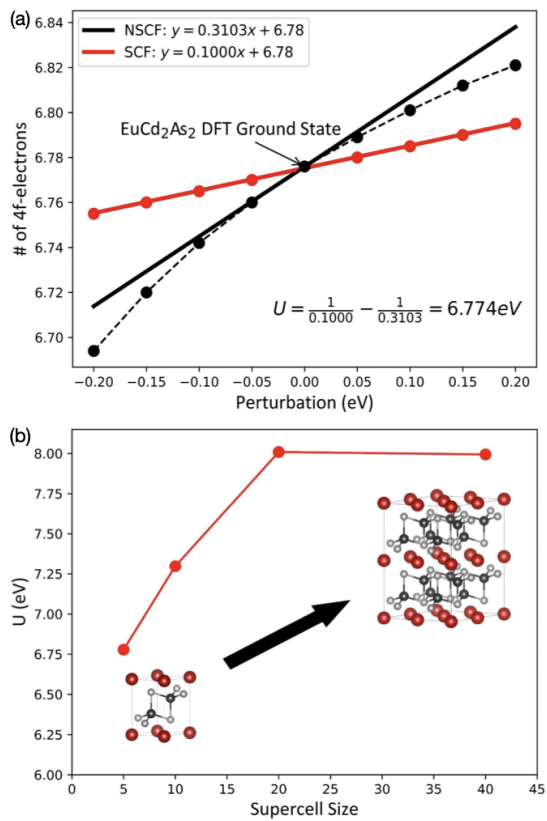} 
\caption{\textbf{First-principle linear response calculations for the Hubbard \textbf{U} of Eu atoms in EuCd$_2$As$_2$.}
\textbf{(a)} Fitted charge self-consistent (solid red line) and non-self-consistent (solid black line) linear response functions for a 5-atom EuCd$_2$As$_2$ unit cell. The inverse slopes of the response functions in the linear regime can be utilized to determine the \textbf{U} value. \textbf{(b)} Convergence test of the calculated Hubbard \textbf{U} parameter with respect to the supercell size. The results converge to \textbf{U} $\sim$ 8 eV in a 40-atom EuCd$_2$As$_2$ unit cell. Such a strong \textbf{U} $\sim$ 8 eV would lead to a finite energy gap and suppression of the topological behavior in EuCd$_2$As$_2$.}
\label{fig6}
\end{figure}

Here, we employ the linear response ansatz by Cococcioni and de Gironcoli to determine the Hubbard \textbf{U} from first principles\cite{PhysRevB.71.035105}. In this approach, the \textbf{U} value is derived by measuring the change in Eu $f$ orbital occupation as a response to external perturbation of an on-site potential. Specifically, after an initial DFT ground state calculation, self-consistent response $\chi$ and non-self-consistent response $\chi_{0}$ are constructed by computing the orbital occupation change respectively with and without DFT charge density updates. Afterwards, \textbf{U} is determined by subtracting the inverse of the self-consistent response from the non-self-consistent one: $\textbf{U}  = \chi^{-1}-\chi^{-1}_{0}$. Figure \ref{fig6}(a) shows the corresponding calculation on a five-atom EuCd$_2$As$_2$ unit cell, in the external potential range -0.20 eV to 0.20 eV. When the perturbation is small, both response functions show a linear trend, so $\chi^{-1}$ and $\chi_{0}^{-1}$ can be obtained directly from the linear slopes. With a larger perturbation, the response functions may exist a non-linear behavior. In this case, the linear response approach can still be applied, by using a non-linear fit and then taking derivative near the zero potential\cite{PhysRevResearch.5.013160}. The result is essentially equivalent to that of a linear fit in a smaller perturbation regime.

As seen in Fig. \ref{fig6}(a), the Hubbard \textbf{U} for a five-atom EuCd$_2$As$_2$ unit cell is calculated to be 6.77 eV. Since the \textbf{U} value is largely affected by charge screening, it is necessary to examine the convergence in larger DFT supercells. We have performed corresponding linear-response \textbf{U} calculations up to a unit cell of 40 atoms, where the \textbf{U} value is found to converge to $\sim$ 8 eV, as shown in Fig. \ref{fig6}(b). We also performed the calculations using different DFT ground states – including paramagnetic and AFM – and found that the resulting \textbf{U} values only differ by less than $\sim$ 0.1 eV. Unlike the empirical values of \textbf{U} $\sim$ 3-5 eV considered in the literatures\cite{10_induced_WSM,8_type-IV_magnetic_space_groups,11_Dirac_AFM_ECA,PhysRevB.98.125110,6_magnetic_order}, the strong Hubbard repulsion \textbf{U} $\sim$ 8 eV discovered in our \textit{ab initio} calculations thereby suggests a finite energy gap and only trivial band topology in EuCd$_2$As$_2$. We also note that a meta-GGA or hybrid DFT functional would lead to an ever larger energy gap that further suppresses the topological behavior\cite{PhysRevB.108.075150}. Therefore, in accord with our carrier-control transport experiments, the theoretical results also indicate that EuCd$_2$As$_2$ is likely to be a topologically trivial magnetic semiconductor.

\section{\label{sec:level1}Conclusion}
Our study revealed that the carrier density dependence of the anomalous Hall conductivity (AHC) in EuCd$_2$As$_2$ does not align with the expected behavior of an ideal Weyl semimetal. Instead, the scaling relationship between AHC and longitudinal conductivity is in good agreement with the characteristics of insulating hopping transport. These observations, coupled with the substantial electronic gap revealed in optical transmittance measurement, unambiguously establish that EuCd$_2$As$_2$ is a magnetic semiconductor instead of a Weyl semimetal. We further performed \textit{ab initio} calculations and discovered that the underestimation of Hubbard \textbf{U} could play a key role in previous misidentification of the topological phase in this material system. Despite being topological trivial, the resistivity of EuCd$_2$As$_2$ is extremely sensitive to the magnetic configuration of Eu-ordered moments, leading to a colossal magnetoresistance effect, indicative of a strong coupling between the electronic band gap and the magnetic orders. This unusual property may be useful for applications in spintronics and magneto-optoelectronics. 

\section{\label{sec:level1}Methods}
\subsection{\label{sec:level2}Crystal synthesis}

EuCd$_2$As$_2$ single crystals were synthesized via the Sn flux method\cite{13_Manipulate_magnetic_inECA}. The stoichiometric Eu, Cd, and As shots with 10, 15, 20, and 25 times atomic mass Sn shots were ground, mixed, and loaded in the bottom of an Canfield crucible\cite{doi:10.1080/14786435.2015.1122248}. Then the crucible was loaded in a quartz tube. To avoid oxidization, grind and load were conducted in the glovebox filled with argon, and the ball valve was used to keep input materials from air when taking it out of the glovebox. The mixture was sealed in an evacuated quartz tube and placed in a box furnace. The mixture was heated to 900 $^{\circ}$C over 12 h , held for 24 h, then slowly cooled over 200 h to 550 $^{\circ}$C. The tube was removed from the furnace and spun in a centrifuge to remove the Sn flux. This process yields plate-like single crystals with hexagonal shape, and size up to a few millimeters. We discovered that the residual carrier density of crystals is highly sensitive to the ratio of Sn flux to start materials. The stoichiometric Eu pieces, Cd and As shots are mixed with 10 to 25 times atomic mass Sn shots as starting materials, which yielded EuCd$_2$As$_2$ crystals with hole carrier density spanning from $10^{15}$ to $10^{16}$ $\text{cm}^{-3}$. This allows us to use various carrier density as a convenient knob to study the electronic band structure of EuCd$_2$As$_2$.

\subsection{\label{sec:level2}Materials characterization}

The composition of all batches of crystals was determined by elemental analysis on a clean crystalline surface using a Sirion XL30 scanning electron microscope. All batches of crystals exhibited nearly perfect stoichiometric composition. A subset of crystals was crushed into fine powder and characterized by powder X-ray diffraction using Rigaku MiniFlex at room temperature. The obtained crystal structure and lattice constants of all batches of crystals are identical to previous reports\cite{7_spectroscopic_of_EuCd2X2}.

\subsection{\label{sec:level2}Transport measurement}

For transport measurements, the sample was cut into a rectangular plate using the wire saw and then made into a standard six-probe contact configuration with the current direction in-plane and the magnetic field out-of-plane ($c$-axis). The contact resistance of samples was limited to few hundreds $\Omega$, which was achieved by sputtering gold pads on sample surface and adhering gold wires with sliver paste. The magneto-transport measurements were carried out in a 9T Physical Property Measurement System (PPMS). We utilized a Keithley2450 as the current source and voltmeter for measuring the $\text{G}\Omega$ level resistance benefiting from the high impedance resistance of Keithley2450. Even though, the current source and measured voltage range should be adjusted accordingly to get a better signal as the resistivity changes a few orders of magnitude as temperature decreases in a short time. To eliminate any effects from contact misalignment, the magneto- and Hall resistance were symmetrized and anti-symmetrized respectively. Magnetization measurements were made using the vibrating sample magnetometry option of the PPMS.

\subsection{\label{sec:level2}Optical measurement}

Fourier-transform infrared (FTIR) spectroscopy was conducted using the Hyperion 2000 system from Bruker Inc. The measurement was performed in the transmission mode, employing an infrared (IR) globar source and a Liquid Nitrogen-cooled Mercury Cadmium Telluride (MCT) detector. To ensure optimal signal transmission, a thin crystal with the flat and clean surface was utilized. Additionally, a reference measurement was conducted using a silicon wafer. 

\subsection{\label{sec:level2}Optical measurement}

Fourier-transform infrared (FTIR) spectroscopy was conducted using the Hyperion 2000 system from Bruker Inc. The measurement was performed in the transmission mode, employing an infrared (IR) globar source and a Liquid Nitrogen-cooled Mercury Cadmium Telluride (MCT) detector. To ensure optimal signal transmission, a thin crystal with the flat and clean surface was utilized. Additionally, a reference measurement was conducted using a silicon wafer.

\subsection{\label{sec:level2}DFT calculation}

Density functional theory (DFT) calculations were performed based on the planewave pseudopotential software VASP (the Vienna Ab initio Simulation Package)\cite{PhysRevB.54.11169,Efficiencyofab-initio}. The calculations utilized the Perdew-Burke-Ernzerhof generalized gradient approximation (PBE-GGA) functional\cite{PhysRevLett.78.1396} and the projector augmented wave (PAW) basis\cite{PhysRevB.50.17953, PhysRevB.59.1758}. Spin-orbit coupling (SOC) was also included. A theoretically relaxed EuCd$_2$As$_2$ crystal structure was considered, with the lattice parameters $a=4.45$ $\text{\AA}$ and $c=7.35$ $\text{\AA}$, which are within 1\% errors compared to the experimental cell\cite{PhysRevB.104.155124, PhysRevB.94.045112}. Convergence tests with respect to the planewave cutoff energy and k-grid size were conducted. The calculations were converged with a cutoff energy of 600 eV, using the global break condition of $10^{-6}$ eV in the electronic self-consistent loops. For the largest 40-atom unit cell under study, a $9\times9\times5$ k-grid was adopted, which corresponds to a fine KPPRA (k-points per reciprocal atom) of 16,200. The DFT ground states provide the input charge densities and wavefunctions for subsequent \textsl{ab initio} linear-response \textbf{U} calculations\cite{PhysRevB.71.035105} to determine the Hubbard \textbf{U} value on Eu atoms.

\section{\label{sec:level1}Data Availability}
Source data are available for this paper. All other data that support the plots within this paper and other findings of this study are available from the corresponding author upon reasonable request.

\begin{acknowledgments}
This work was mainly supported by the Air Force Office of Scientific Research (AFOSR) under Award No. FA2386-21-1-4060. The transport experiment was partially supported as part of Programmable Quantum Materials, an Energy Frontier Research Center funded by the U.S. Department of Energy (DOE), Office of Science, Basic Energy Sciences (BES), under Award No. DE-SC0019443. The FTIR measurement was supported by NSF MRSEC at UW (DMR-2308979). J.-H.C. also acknowledges support from the David and Lucile Packard Foundation and the support from the State of Washington funded Clean Energy Institute.  A portion of this work was performed at the National High Magnetic Field Laboratory, which is supported by the National Science Foundation (NSF) Cooperative Agreement No. DMR-1644779 and the State of Florida.
The calculations were performed on the Frontera computing system at
the Texas Advanced Computing Center made possible by the NSF Award No. OAC-1818253.
\end{acknowledgments}

\appendix
\section{}
\subsection{\label{app:subsec}Refinement of X-ray diffraction}

\begin{figure}[h] 
\includegraphics[width = 3.4 in]{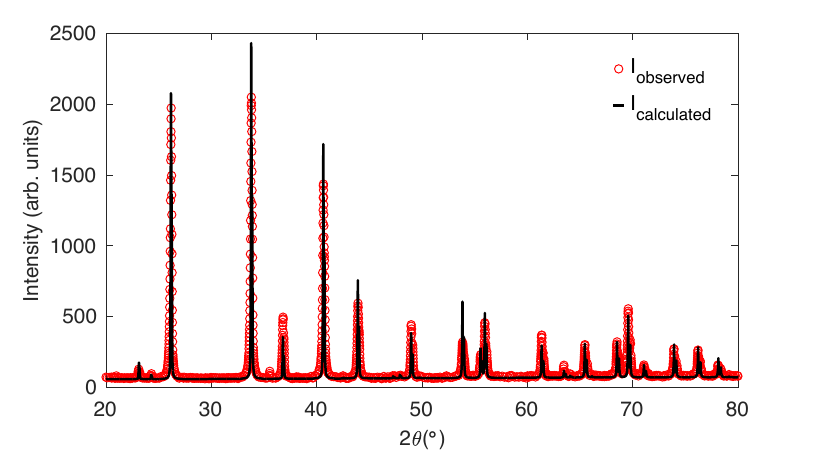}
\caption{\textbf{Refinement of powder X-ray diffraction of EuCd$_2$As$_2$.} The observed X-ray diffraction spectrum and the calculated intensity gained by the Rietveld refinement.}
\label{figs1}
\end{figure}

The X-ray diffraction pattern was obtained from a Rigaku MiniFlex diffrectometer as described in the method section. The observed X-ray diffraction intensity pattern of crystals from 10Sn EuCd$_2$As$_2$ are shown in Fig. \ref{figs1}, along with the calculated intensity made via the Rietveld refinement of observed intensity using FullProf suite\cite{RODRIGUEZCARVAJAL199355}. The crystal lattice constants from refinement are $a = 4.4386$ \AA and $c = 7.3239$ \AA, which agree with the previous reports\cite{6_magnetic_order,13_Manipulate_magnetic_inECA,PhysRevB.94.045112}.

\vspace{10pt}

\subsection{\label{app:subsec}High-field Magnetoresistance}

\begin{figure}[h]
\includegraphics[width = 3.4 in]{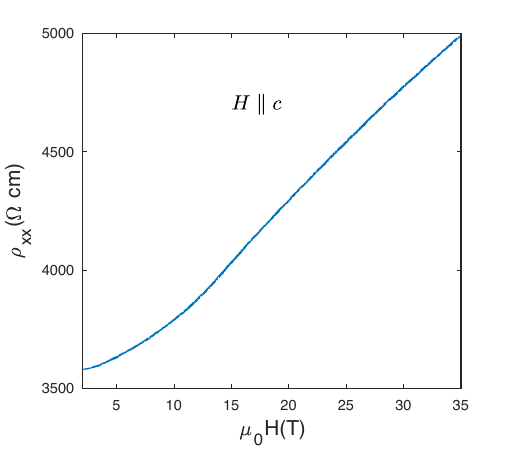} 
\caption{\textbf{High-field magnetoresistance.}
The magnetoresistivity $\rho_{xx}$ versus along $c$-axis magnetic field $\mu_{0}H$ range from 2 T to 35 T of the sample from 10Sn EuCd$_2$As$_2$.}
\label{figs2}
\end{figure}

High-field magnetoresistance measurements were conducted under up to 35 T magnetic field at National High Field Magnetic National Laboratory (NHMFL) in Tallahassee, Florida. Figure \ref{figs2} shows the $\rho_{xx}$ measured on a sample from 10Sn EuCd$_2$As$_2$ with magnetic field along $c$-axis at 1.4 K. The magneto-resistivity presents a quadratic increase as field increases. There is no Shubnikov-de Haas (SdH) quantum oscillations appears until 35 T.

\subsection{\label{app:subsec}Magnetoresistance of samples from 15Sn and 20Sn}

\begin{figure}[h]
\includegraphics[width = 3.4 in]{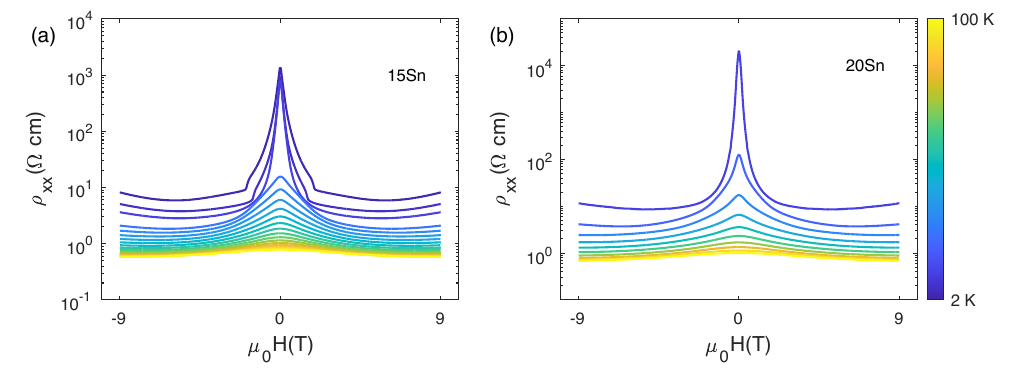} 
\caption{\textbf{Colossal Magnetoresistance.}
Temperature dependence of resistivity, $\rho_{xx}$, versus $\pm$9 T magnetic field along $c$-axis at temperatures between 2 K and 150 K for samples from \textbf{(a)} 15Sn, and \textbf{(b)} 20Sn.}
\label{figs3}
\end{figure}

The magneto-resistivity $\rho_{xx}$ as a function of field were also conducted on samples from 15Sn and 20Sn. Fig. \ref{figs3} shows $\rho_{xx}$ versus magnetic field at fixed temperature between 2-150 K. The behaviors are qualitatively same as the represented data of sample from 10Sn shown in the main content. 

\bibliography{ECA}

\end{document}